\documentclass[%
 aps
 ,
 sd,%
 amsmath,amssymb,reprint,
jmp]{revtex4-1}%
\usepackage{graphicx}% Include figure files
\usepackage{dcolumn}% Align table columns on decimal point
\usepackage{bm}% bold math
\usepackage{color,soul}
\usepackage{comment}
\usepackage{float}
\usepackage[section]{placeins}
\newcommand{\vect}[1]{\bm{#1}}

\begin{document}

\title{Transient dynamics of nonlinear magneto-optical rotation in the presence of transverse magnetic field  }
\author{Raghwinder Singh Grewal}
\email{raghugrewal.singh@gmail.com}
\author{Szymon Pustelny}
\email{pustelny@uj.edu.pl}
\affiliation{Institute of Physics, Jagiellonian University, \L ojasiewicza 11, 30-348 Krak\'ow, Poland}
\begin{abstract}

Nonlinear magneto-optical rotation is studied under non-equilibrium conditions. The polarization rotation of linearly polarized light traversing a rubidium vapor cell is observed versus the time-dependent (swept) longitudinal magnetic field in the presence of static transverse  magnetic fields. Presence of the transverse fields modifies the character of the observed signals. In particular, for weaker transverse fields, field sweep leads two-harmonic oscillation of the polarization rotation while crossing zero. Unlike the steady-state, it was found that two-frequency oscillations observed in the transient signals, are  independent of the transverse-field direction. For stronger transverse fields, the oscillations deteriorate eventually reaching a situation when no-oscillating dynamic signal, with distinct minimum close to zero field, is observed. Experimental results are supported with theoretical analysis based on the density-matrix formalism. The analysis confirms all the features of experimental results while providing an provide intuitive explanation of the observed behavior based on angular-momentum probability surfaces used for density-matrix visualization.

\end{abstract}

\maketitle

\section{Introduction}

Coherent interaction of  resonant light with atoms results in an onset of many interesting optical phenomena including electromagnetically induced transparency \citep{Fleischhauer}, electromagnetically induced absorption \citep{LezamaEIA,RaghwinderEIA}, coherent population trapping \citep{NasyrovCPT,RaghwinderCPT}, and nonlinear magnetic-optical rotation (NMOR) \citep{Budker1}. These effects are usually studied when an atomic system is in a steady state, where dynamic equilibrium between various processes is achieved. A specific example of such studies is NMOR \citep{Budker0,Budker1,Budker2,Malakyan,Novikova:05,Zigdon,Pustelny2015Nonlinear}, a process of rotation of linear polarization of light traversing an optically-polarized medium subjected to a magnetic field \citep{Budker1}. At a microscopic level, NMOR can be explained as (1) initial optical pumping of atoms into an aligned state (generation of optical anisotropy in atoms) and (2) successive evolution of the state due to magnetic field (precession of the anisotropy) or other parameters, leading to further modifications of the state (e.g., anisotropy deterioration). Assuming time independent parameters of the system (e.g., magnetic field, relaxation rates etc.), under continuous illumination, a dynamic equilibrium between the processes is reached and a net static optical anisotropy in the medium is generated. In contrast to the steady-state case, NMOR under non-equilibrium situation, i.e., when experimental conditions change faster than the characteristic time scale of a system is studied less frequently \citep{Momeen,RaghwinderNMOR,Jin:19}. Under such conditions, the contribution from various processes to the overall anisotropy of the medium leads to complex time-dependent dynamics of the system, resulting in signals that differ significantly from those observed in conventional NMOR experiments. 

Recently, theoretical and experimental studies of NMOR with time-dependent magnetic fields were performed \citep{RaghwinderNMOR}. In the studies, the polarization rotation of continuous-wave (CW) light, illuminating rubidium vapors, was studied versus light (intensity) and magnetic-field parameters (amplitude, frequency, and sweep rate).  The studies demonstrated that at low sweep rates, a traditional dispersive NMOR signal, with a small asymmetry between its two sides, is observed. It was also shown that for faster sweep rates, the damped oscillations of the rotation signal are recorded. It was found that this behavior strongly depends on magnetic-field sweep rate, but not on amplitude or frequency of the scan (same behavior was observed for different magnetic-field amplitudes and frequencies but same sweep rates). 

The aim of the present work is to study NMOR as a function of a time-dependent longitudinal magnetic field $B_z(t)$. In contrast to the previous work \citep{RaghwinderNMOR}, however, the measurements are performed in the presence of the static magnetic field $B_x$ or $B_y$, oriented transversely to the light-propagation direction. Presence of transverse field modifies evolution of the system, resulting in changes of NMOR signals. In particular, an additional modulation in the signals is observed when a weak transverse field is applied.  Next, it is demonstrated that for stronger transverse fields the signals change their character. Under such conditions, not only the oscillations are suppressed, but also the shape of the signal envelop changes, it becomes absorptive rather than dispersive. This behavior is observed experimentally and is confirmed with numerical simulations based on the density-matrix formalism. Visualization of the density-matrix, using angular-momentum probability surfaces, provides intuitive understanding of the observed process, keeping all rigorousness of the quantum-mechanical calculations.
 
\section{Theoretical Results and discussion\label{sec:Results}}

In our analysis, NMOR signals are calculated using the density-matrix formalism. While details of the calculations can be found elsewhere \citep{RaghwinderNMOR}, here we just recall their key elements. The analysis is performed in a $F=2\rightarrow F'=1$ atomic system. The atoms interact with $y$-polarized light, propagating along the quantization axis $z$, with the wavevector $k_z$ and of the amplitude $E_y$. The light-atom interaction is considered within the dipole approximation, in which interaction of light results in repopulation of ground-state magnetic sublevels and generation of Zeeman coherences in the states. The time-dependent magnetic field $B_{z}(t)$ is applied along the quantization axis, leading to energy splitting of Zeeman sublevels. Here, we only consider a linear Zeeman effect due to a small value of the applied field. Simultaneously, the static fields $B_x$ and $B_y$ are generated in transverse directions $x$ and $y$, respectively, coupling the magnetic sublevels. In the analysis, two types of relaxation are assumed. The first originates from spontaneous emission (density-matrix dependent relaxation) and the second is uniform relaxation, repopulating atoms toward the thermal equilibrium (density-matrix-independent relaxation). The time-dependent evolution of the density matrix is calculated within the rotating-wave approximation by numerically solving the Liouville equation. Finally, the density matrix is used to calculate an angle of polarization rotation \citep{AuzinshOptically}. While this model does not fully reproduce the experimental system, where atoms travel freely between bright (illuminated) and dark regions of the cell, and more complex energy structure (e.g., hyperfine levels) is present, it still captures all essential elements required for the simulations. 

\begin{figure}[H]
\centering
\includegraphics[height=5.5cm,width=7.5cm]{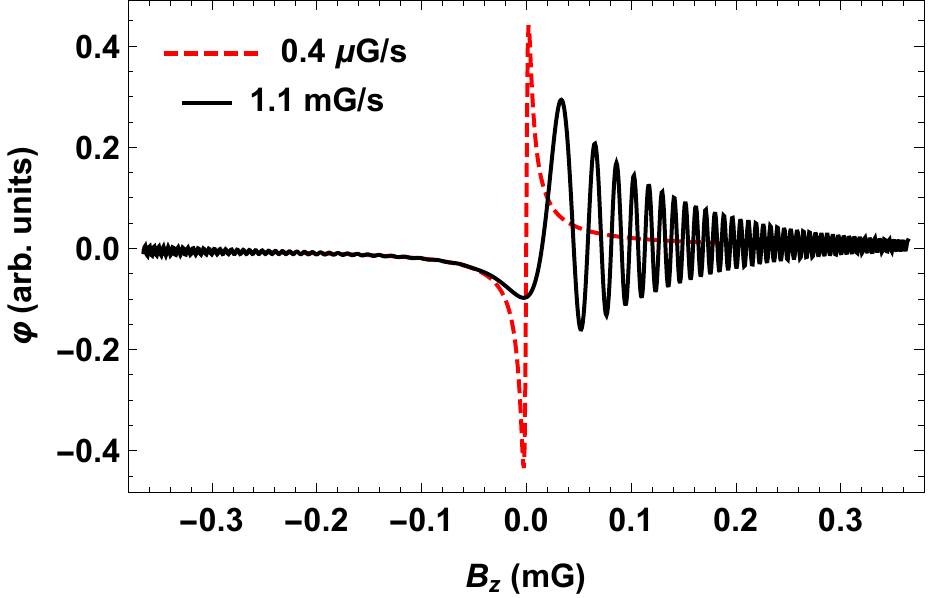}	
\caption{Numerically simulated NMOR signals as a function of the longitudinal magnetic field $B_z(t)$ at different sweep rates (0.4~$\mu$G/s and 1.1~mG/s) and no transverse fields ($B_x=B_y=0$). Other parameters used in simulations  are: the optical Rabi frequency ${\Omega _R}/2\pi =0.01\,{\rm{MHz}}$, the ground-state relaxation $\gamma/2\pi = 2.5\,{\rm{Hz}}$, and the excited-state relaxation of $\Gamma/2\pi = 5.75\,{\rm{MHz}}$. The oscillations seen on the left-hand side of the plot is an artifact of our simulation which are performed for continuously sweep field with a triangular function \citep{RaghwinderNMOR}.}
	\label{diffSweepRate}
\end{figure} 

Figure~\ref{diffSweepRate} presents numerically calculated NMOR signals at different magnetic-field sweep rates without any transverse fields. The plots show that at a low sweep rate, a dispersively shaped signal (red dashed curve), corresponding to a conventional (steady state) NMOR experiment, is observed. When the sweep rate increases, the chirped damped oscillations are observed while crossing zero field (black solid curve). This indicates that the medium does not reach equilibrium state during the sweep. The chirping observed in the signal originates from temporal variation of the magnetic field, while the damping is a result of washing out medium optical anisotropy through combined action of magnetic field and CW optical pumping.

\begin{figure}[H]
\centering  
\includegraphics[height=11cm,width=8.0 cm]{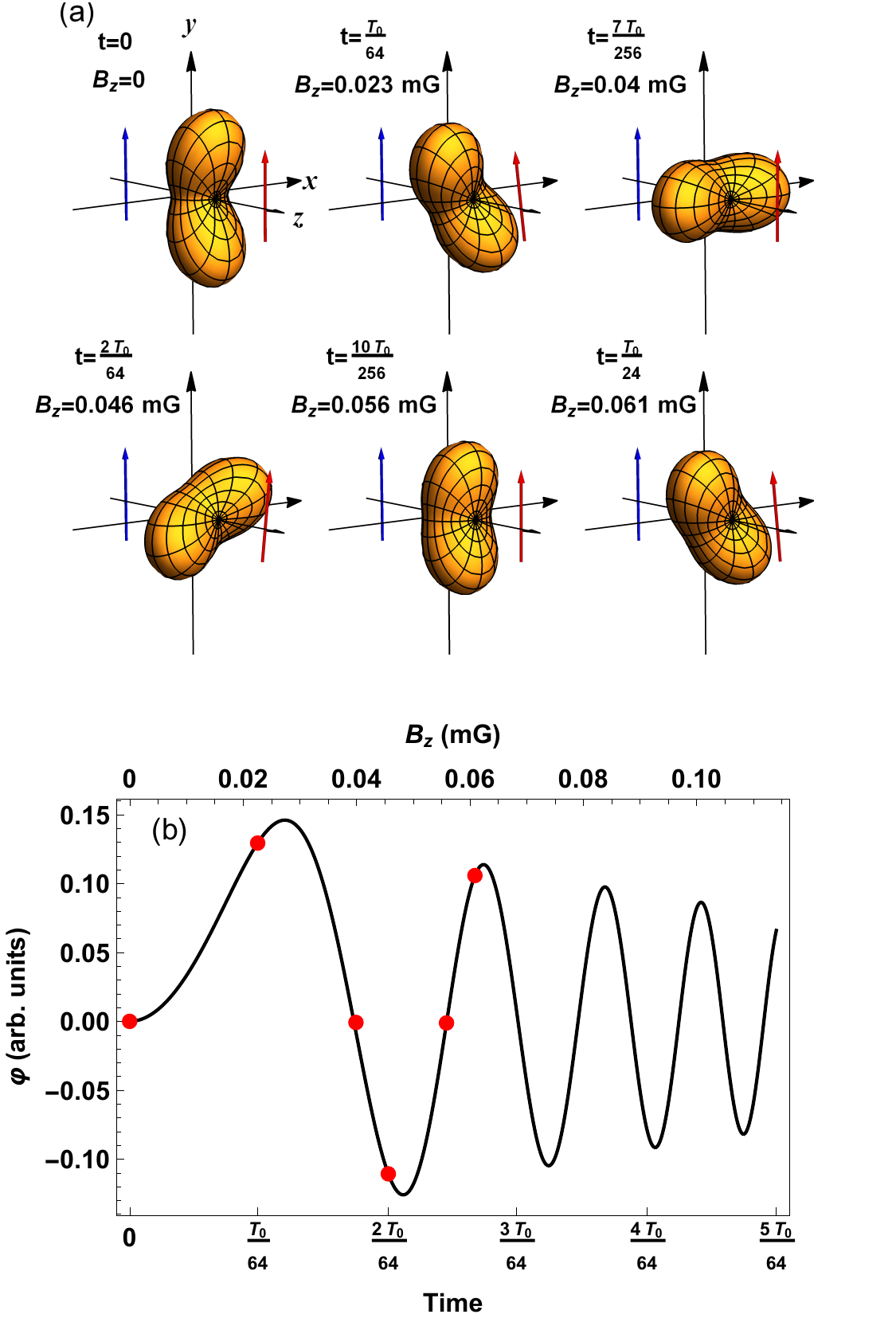}
\caption{(a) Angular-momentum probability surfaces illustrating temporal evolution of the density matrix. For better clarity, at time $t<0$ only interaction of strong pump light with atoms was considered in the absence of any magnetic-field. This led to a strong atomic polarization shown in the first plot ($t = 0$). Successive evolution of the system with time-dependent magnetic field and much weaker probe light was next considered ($t>0$). (b) Corresponding optical rotation signals as function of time while scanning the longitudinal magnetic field (note differences with respect to Fig.~\ref{diffSweepRate}, e.g., at $t=0$, originating from separated optical pumping and magnetic precession). Simulations are performed for the same set of parameters as in Fig.~\ref{diffSweepRate}, including a sweep rate of 1.1~mG/s. $T_{0}$= $B_0$/$S$ is the time period, where $B_0$ is the amplitude of the field sweep and $S$ is the sweep rate. Blue and red arrows in (a) show the polarization rotation angles before and after the medium. Red points marked on polarization-rotation curve in (b) are positioned  corresponding to the probability surface plots in (a). }
\label{ProwithoutTMF}
\end{figure}
To investigate the quantum-mechanical evolution of the system in more details, we present the density matrix at various stages of evolution, in a form of angular-momentum probability surfaces \citep{Rochester2001}. The surfaces help in getting a clear and intuitive understanding of various features of the system, including spatial symmetries of the density matrix. These symmetries help to deduce specific features of the associated optical anisotropy and hence determine properties of the NMOR signals.

\begin{figure}[H]
\centering
\includegraphics[height=8.3cm,width=\columnwidth]{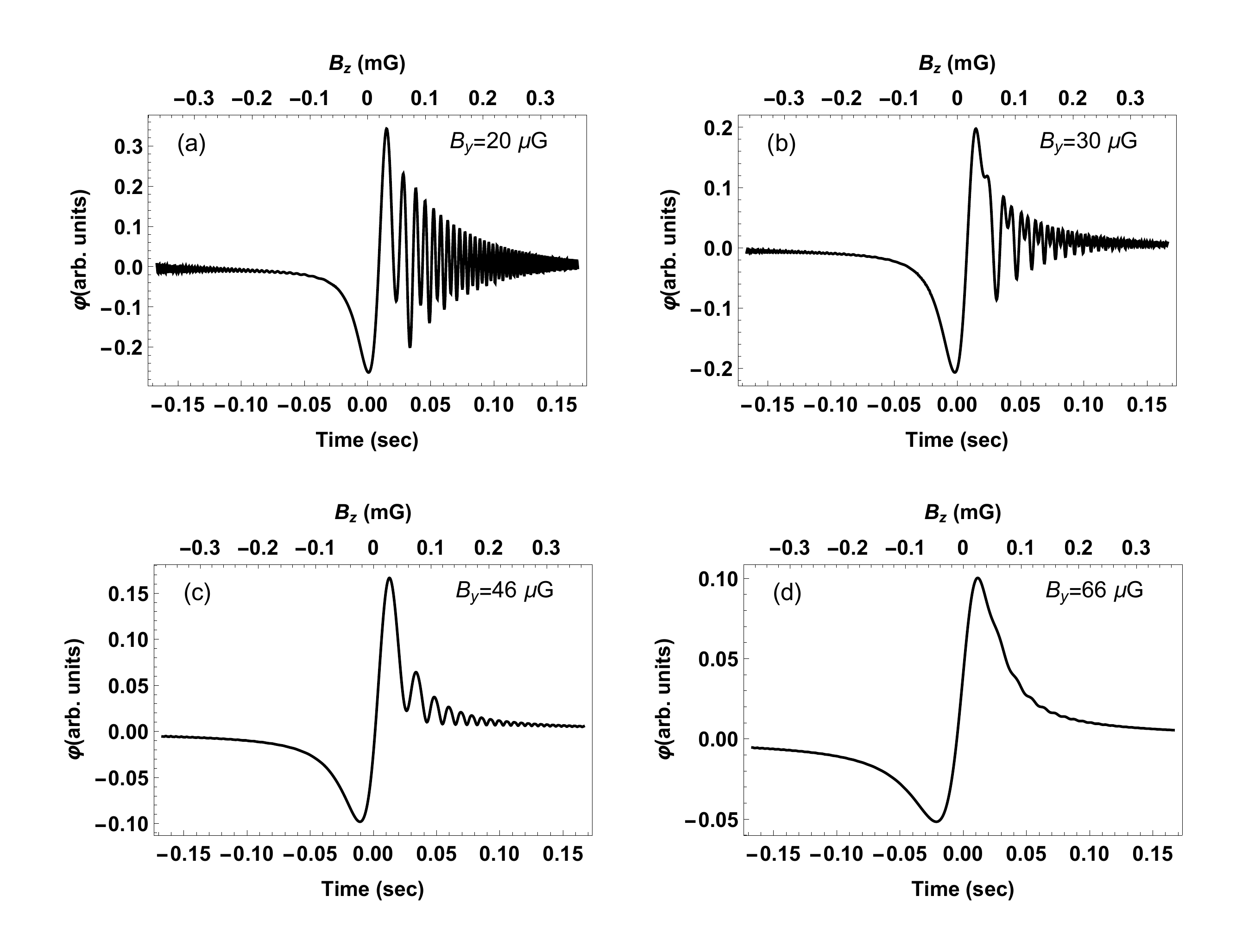}
\caption{NMOR signals simulated as a function of time/longitudinal magnetic field at a sweep rate of $1.1$~mG/s and different $B_y$ fields: (a)  $B_y = 20$~$\mu$G, (b) $B_y=30$~$\mu$G, (c) $B_y=46$~$\mu$G, and (d) $B_y=66$~$\mu$G. All calculations were performed for $B_x=0$ and other parameters of the simulations are the same as in Fig~\ref{diffSweepRate}.}
\label{diffTMFY}
\end{figure}
For better understanding of the observed dependences, we consider a simplified situation, where atoms are first optically pumped with strong light at $\vect{B_z}=0$ ($t\leq 0$), so that an atomic alignment is created. Then the time evolution of the state is investigated when a weak probe beam is present and the longitudinal magnetic field is scanned with nulled transverse fields ($B_x=B_y=0$). Figure~\ref{ProwithoutTMF}(a) presents the probability surfaces corresponding to such evolution. Timings for the probability surfaces plots are chosen in such a way to show proper dynamics of the medium, which corresponds to red dots in polarization rotation curve in Fig.~\ref{ProwithoutTMF}(b). As shown, the pumping of atomic alignment corresponds to creation of a peanut-like-shaped probability surface. Successive Zeeman interaction results in precession of the shape around the magnetic field (Larmor precession). The Larmor frequcny $\Omega_L= \gamma B_z$, where $\gamma$ is the gyromagnetic ratio of the atom. Since the magnetic field $B_z$ changes over time, the precession frequency also increases. Moreover, due to the relaxation and continuous (weak) optical pumping of the atoms the system relaxes into a different state (with no Zeeman coherences), and anisotropy of the whole medium deteriorates. As light experience weaker optical anisotropy, its polarization rotation gets smaller. As a result, the process leads to slow damping of chirped oscillations of the polarization plane presented in Fig.~\ref{ProwithoutTMF}(b).

The situation changes when an additional transverse field is applied to the atoms. First, we considered the case when the transverse magnetic field $B_y$ is applied along the light-polarization direction. In the case, the net magnetic field $\vect{B_{net}}=\vect{B_y}+\vect{B_{z}}(t)$ is oriented in the $yz$ plane, i.e. the plane containing atomic alignment initially induced by light. 

\begin{figure}[H]
\centering  
\includegraphics[height=11.0cm,width=8.0cm]{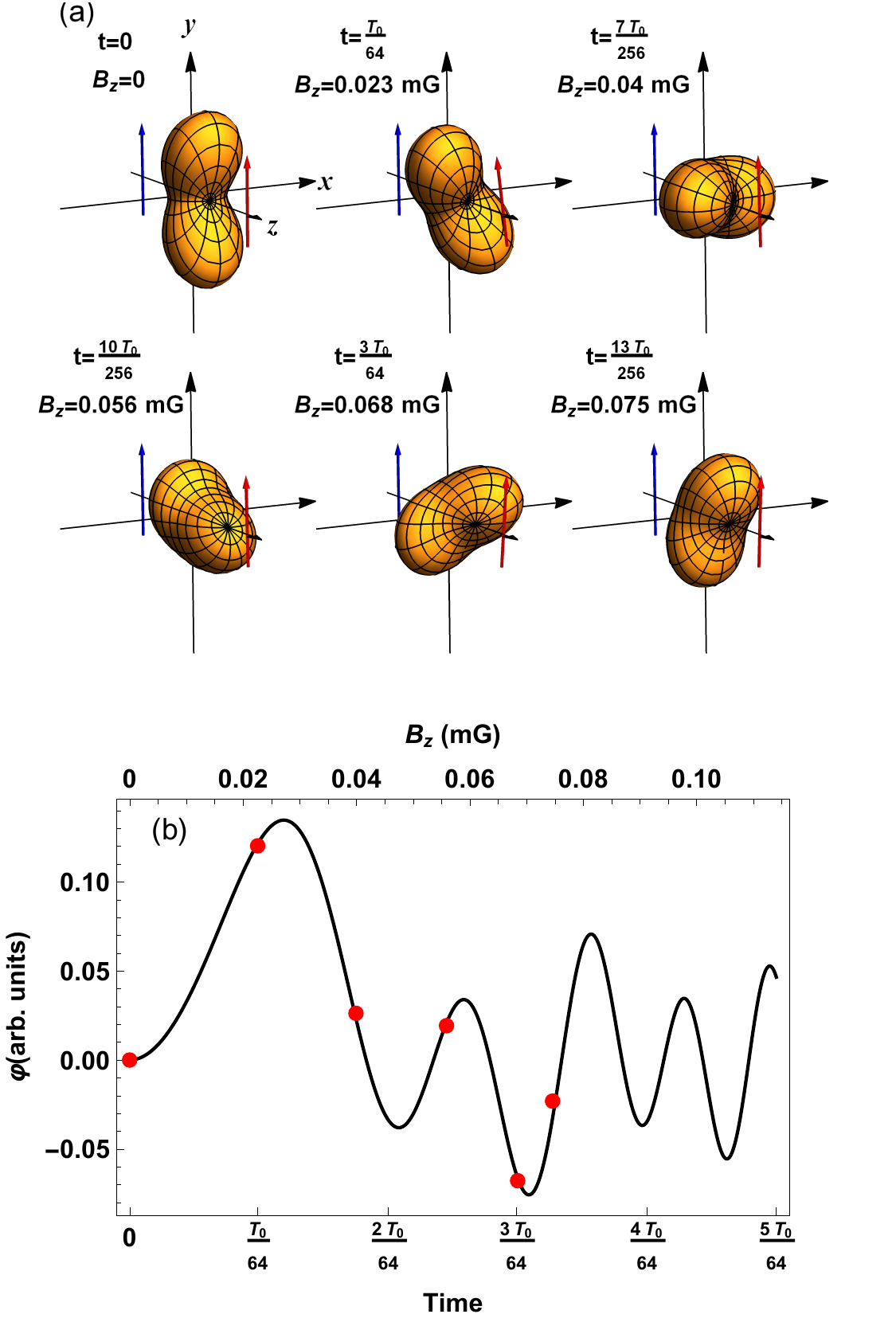}
\caption{(a) Angular-momentum probability surfaces illustrating evolution of the density matrix due to a combine action of time-dependent longitudinal magnetic field and the transverse magnetic field $B_y=20$~$\mu$G ($B_x=0$). (b) Corresponding optical rotation signal as function of time while scanning the longitudinal magnetic field. Other parameters used in simulations are same as in Fig.~\ref{ProwithoutTMF}.}
\label{ProwithBYTMF}
\end{figure}
Figure~\ref{diffTMFY} presents the rotation signals as a function of time at different $B_y$. As shown, for the relatively weak transverse field ($B_y=20$~$\mu$G), a two-frequency ($\Omega_L(t)$ and 2$\Omega_L(t)$) oscillation is observed. In the case, the 2$\Omega_L$ oscillations dominates in the signal. Similar two-frequency oscillations were described in Ref.~\citep{Lenci} and its time-independent (demodulated) counterpart, was investigated in Ref.~\citep{Pustelny}, under the steady state, where two-frequency oscillations manifested as resonances at $\Omega_L$ and $2\Omega_L$.

As $B_y$ increases [Fig.~\ref{diffTMFY}(b)], the amplitude of the 2$\Omega_{L}$ oscillation becomes smaller than the amplitude of the $\Omega_{L}$ oscillation. For stronger fields $B_y\gtrsim 40$~$\mu$G, the 2$\Omega_{L}$-oscillation vanishes completely [Fig.~\ref{diffTMFY}(c)]. For even stronger fields ($B_y\gtrsim 60$~$\mu$G) no oscillations (neither at $\Omega_L$ nor $2\Omega_L$) are observed in the NMOR signal. This dependence on the transverse-field magnitude  corresponds well to the steady-state situation, where amplitude ratio of the $\Omega_L$ and $2\Omega_L$ resonances depend on the orientation of the magnetic field in the $xy$ plane \cite{Pustelny}.

\begin{figure}[H]
\centering  
\includegraphics[height=8.3cm,width=\columnwidth]{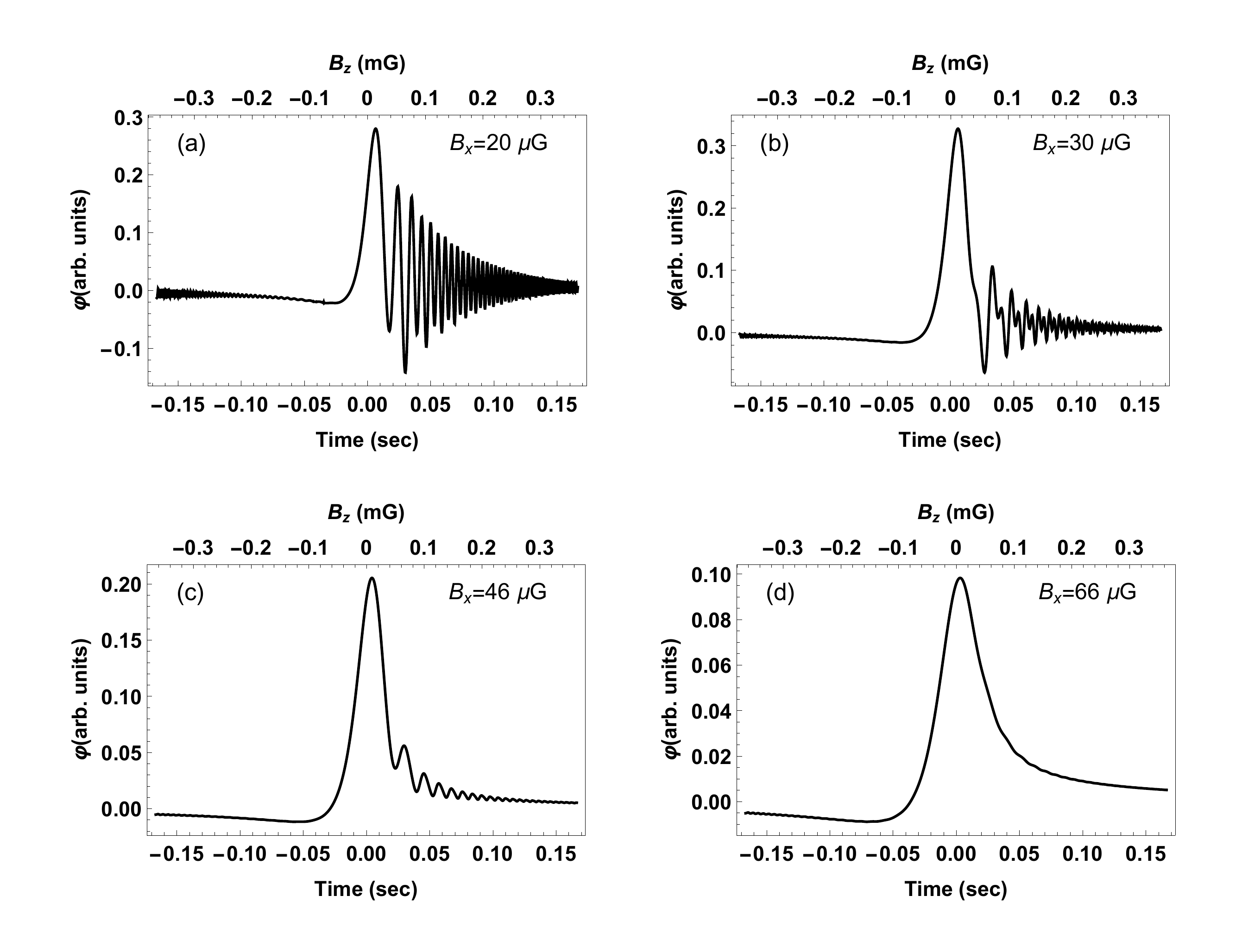}
\caption{Simulated NMOR signals as a function of time/magnetic field at a sweep rate of $1.1$~mG/s at different values of the transverse field $B_x$: (a)  $B_x=20$~$\mu$G, (b) $B_x=30$~$\mu$G, (c) $B_x=46$~$\mu$G, and (d) $B_x=66$~$\mu$G and $B_y=0$. Other simulation parameters are identical as in the previous cases.}
\label{diffTMFX}
\end{figure}
Figure~\ref{ProwithBYTMF}(a) shows the temporal evolution of the density matrix in the presence of $B_{y}$, while scanning the longitudinal field $B_z(t)$. At $t=0$, the longitudinal magnetic field is zero, so the net field is oriented along the light polarization and atomic-alignment direction. Orientation of the magnetic field along the alignment axis does not modify the state of the atoms, and hence no polarization rotation is initially observed. As the longitudinal field $B_z(t)$ increases, the net magnetic field starts changing its direction, tilting in the \textit{yz} plane toward $z$.  Since it is no longer perpendicular to the alignment axis, the alignment starts to precesses around the instantaneous field. After half of the Larmor period, the projection of this alignment onto the $xy$ plane is parallel to the initial orientation, however its length (amplitude of the transverse-alignment projection onto the $xy$ plane) is different from the initial case. As projection of the alignment on the $xy$ plane determines the strength of optical anisotropy of the medium, the anisotropy of the medium is  weaker, so the induced polarization rotation when rotated from the orientation is smaller. When the $B_y$ field is not too strong, the initial state is better reproduced after another half Larmor period, i.e., the full precession period. Due to this, one observes two-frequency modulation of the rotation signal at $\Omega_L(t)$ and $2\Omega_L(t)$ [Fig.~\ref{ProwithBYTMF}(b)].
 On the top of the magnetic evolution, the processes of relaxation and repumping, leading to deterioration of the transverse alignment and damping of the oscillations is observed [Fig.~\ref{ProwithBYTMF}(b)].
\begin{figure}[h]
\centering  
\includegraphics[height=11.0cm,width=8.0cm]{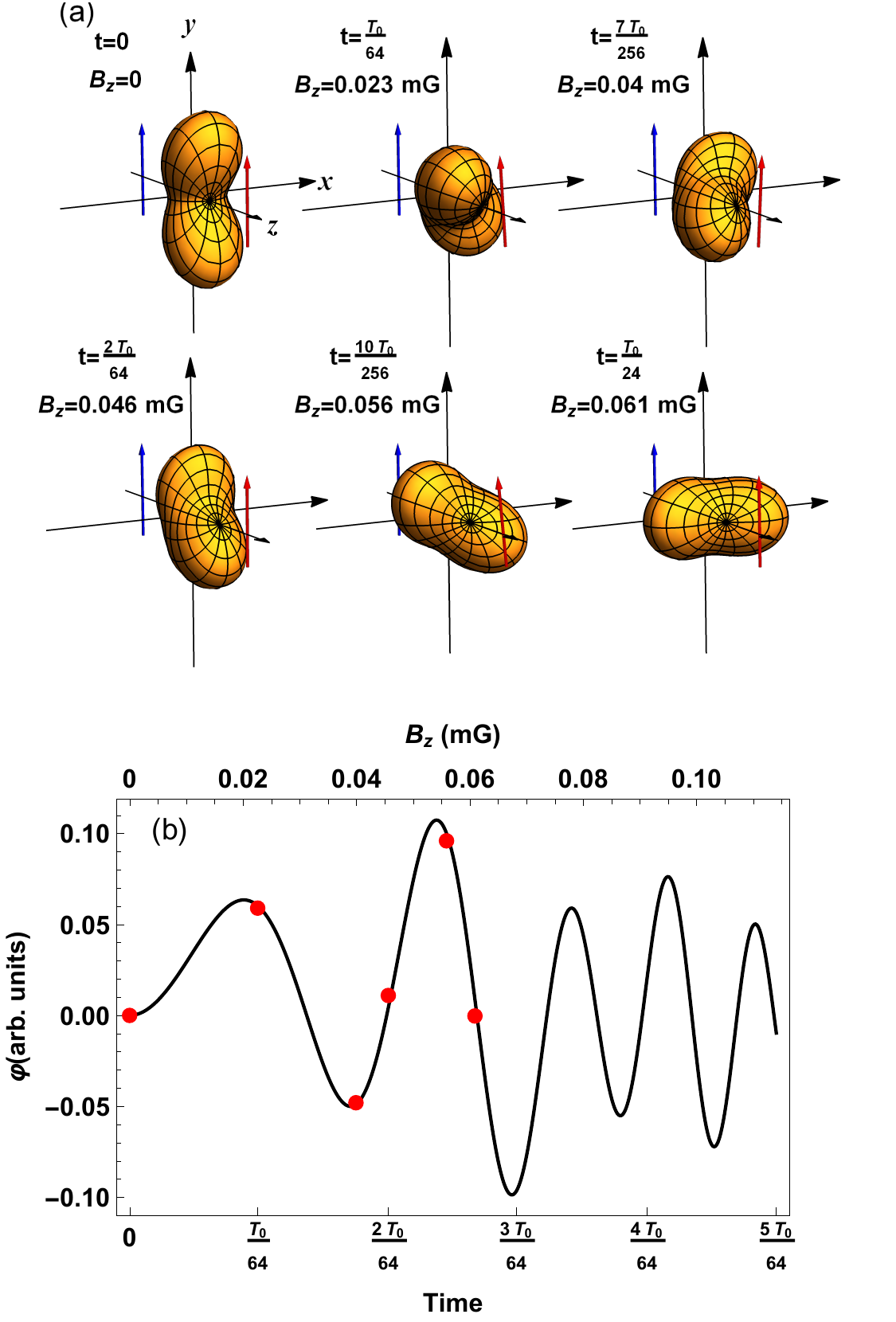}
\caption{(a) Angular-momentum probability surfaces illustrating the time-dependent dynamics of the density matrix and (b) corresponding optical rotation signal. The simulations are performed with $B_{x}=20$~$\mu$G and $B_y=0$, and other simulation parameters unchanged with respect to earlier simulations.}
\label{ProwithBXTMF}
\end{figure}

Next, the polarization rotation is calculated for different $B_{x}$ but $B_y=0$ (Fig.~\ref{diffTMFX}). As demonstrated in Ref.~\cite{Pustelny}, under the steady state, such a constant field orientation leads to significantly different results as no resonance at $\Omega_L$ was observed. However, when the longitudinal field $B_z$ is scanned, the behavior is different, i.e., two-frequency modulation is again observed [Fig.~\ref{diffTMFX}(a)]. In fact, for the weak transverse magnetic fields, the signals in the presence of $B_y$ [Fig.~\ref{diffTMFY}(a)] and $B_x$ [Fig.~\ref{diffTMFX}(a)] are similar (comparable amplitude ratio of the two frequency components). The only distinct feature arises for negative magnetic field; in the case of $B_y>0$~$\mu$G and $B_x=0$, the rotation is negative at $B_z(t)\lesssim 0$~$\mu$G, but for $B_x>0$ and $B_y=0$, this negative rotation is strongly suppressed. The similarity between the two cases becomes less apparent for the stronger transverse fields. This is well visible in Figs.~\ref{diffTMFX}(b) \& (c), where strong differences between envelopes of two signals but also amplitudes of two components [Fig.~\ref{diffTMFX}(b)] are observed. Finally, for the strongest fields ($B_x\gtrsim 60$~$\mu$G), no oscillations are present in the rotation signal [Fig.~\ref{diffTMFX}(d)].
%, but the whole signal has a different (absorptive) shape than for $B_x=0$ [Fig.~\ref{diffTMFX}(d)].

Figure~\ref{ProwithBXTMF}(a) shows the time evolution of the density matrix in the presence of nonzero $B_{x}$ and $B_y=0$. At time $t=0$, the longitudinal magnetic field is zero, so the net magnetic field is directed along the \textit{x}-axis. Since magnetic field is perpendicular to the alignment, it tilts the alignment in the \textit{yz} plane, so that angle between alignment and light propagation direction is smaller than 90$^\circ$. Although this tilt still does not produce any polarization rotation at time $t=0$ [Fig.~\ref{ProwithBXTMF}(b)], but the situation corresponds with the case of $B_y\neq 0$ (it changes the projection of the alignment onto the $xy$ plate). When the longitudinal field increases, the alignment tilted from the $xy$ plane starts to precess around the instantaneous magnetic field, more precisely reproducing itself after full Larmor period rather than after half of the period. This process results in two-frequency components of the time-dependent rotation [Fig.~\ref{ProwithBXTMF}(b)].

\begin{figure}[H]
\centering  
\includegraphics[height=8.3cm,width=\columnwidth]{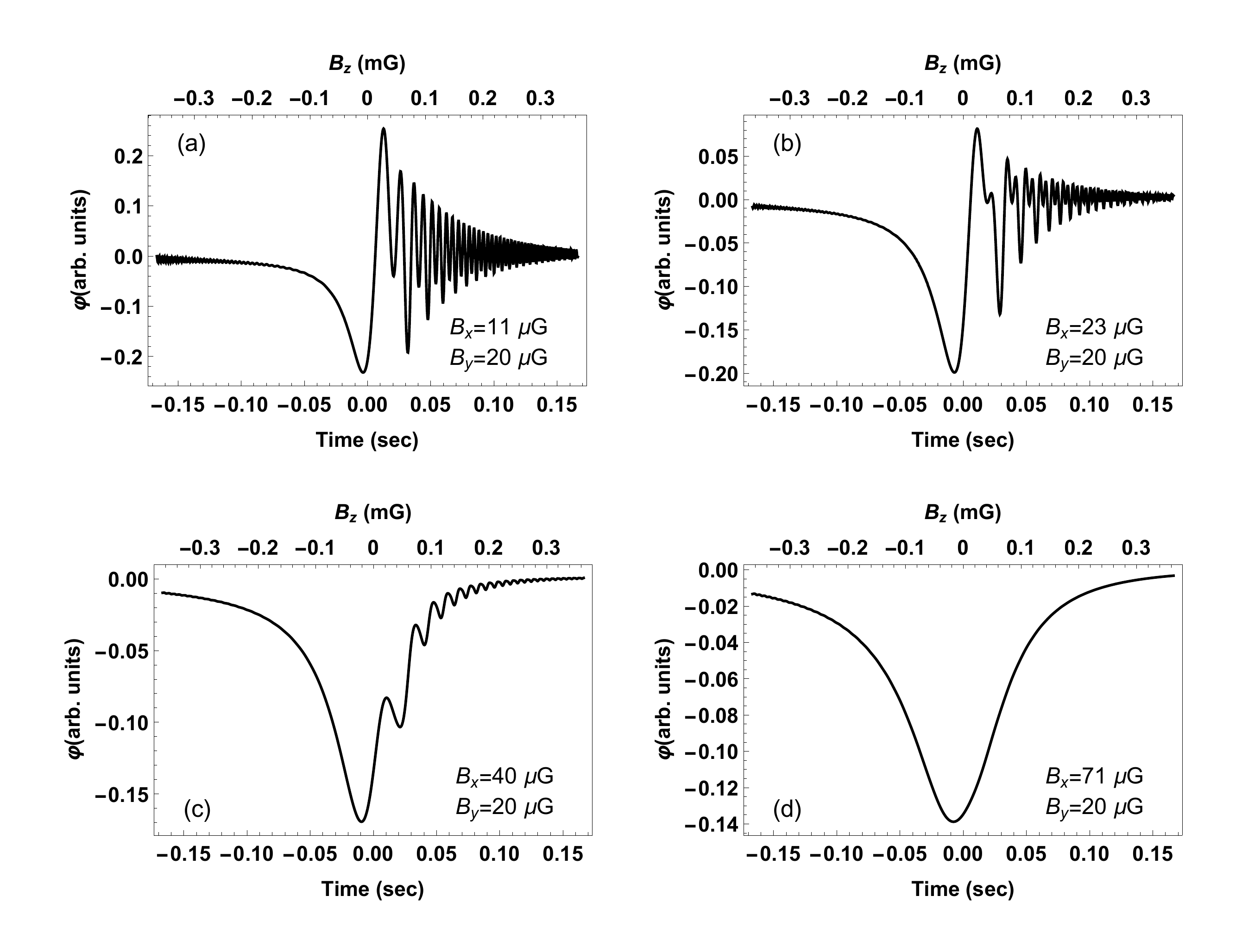}
\caption{Simulated NMOR signals versus time/magnetic field at a sweep rate of $1.1$~mG/s, $B_y=20$~$\mu$G and (a)  $B_x=11$~$\mu$G, (b) $B_x=23$~$\mu$G, (c) $B_x=40$~$\mu$G, and (d) $B_x=71$~$\mu$G. Other parameters are identical as in previous simulations.}
\label{diffTMFxy}
\end{figure}

Figure~\ref{diffTMFxy} presents the polarization rotation signals when the magnetic fields $B_{x}$ and $B_{y}$ are simultaneously nonzero. The signals are calculated at different values of  $B_{x}$  with fixed  $B_{y}$ ($B_{y}$= 20 $\mu$G).  When $B_{x}$ is less than $B_{y}$ [Fig.~\ref{diffTMFxy}(a)], the transient oscillations are similar to that observed in the signals shown in Fig.~\ref{diffTMFY}(a). As $B_{x}$ increases, the signal becomes weaker [Fig.~\ref{diffTMFxy}(b)].
But, the amplitude of the positive rotation angles decreases more compared to negative angles. As $B_{x}$ is further increased, the transient oscillations only with negative rotation angles are seen [Fig.~\ref{diffTMFxy}(c)]. At very high $B_{x}$ ($B_{x}$ $\gg B_{y}$) [Fig.~\ref{diffTMFxy}(d)], the oscillations in the rotation signal vanishes and the shape of the signal changes to absorptive [Fig.~\ref{diffTMFxy}(d)], which is different from the signals in Fig.~\ref{diffTMFY}(d) and \ref{diffTMFX}(d).

\section{Experimental Results and discussion}

\begin{figure}[h]
\centering  
\includegraphics[height=6.0cm,width=9.cm]{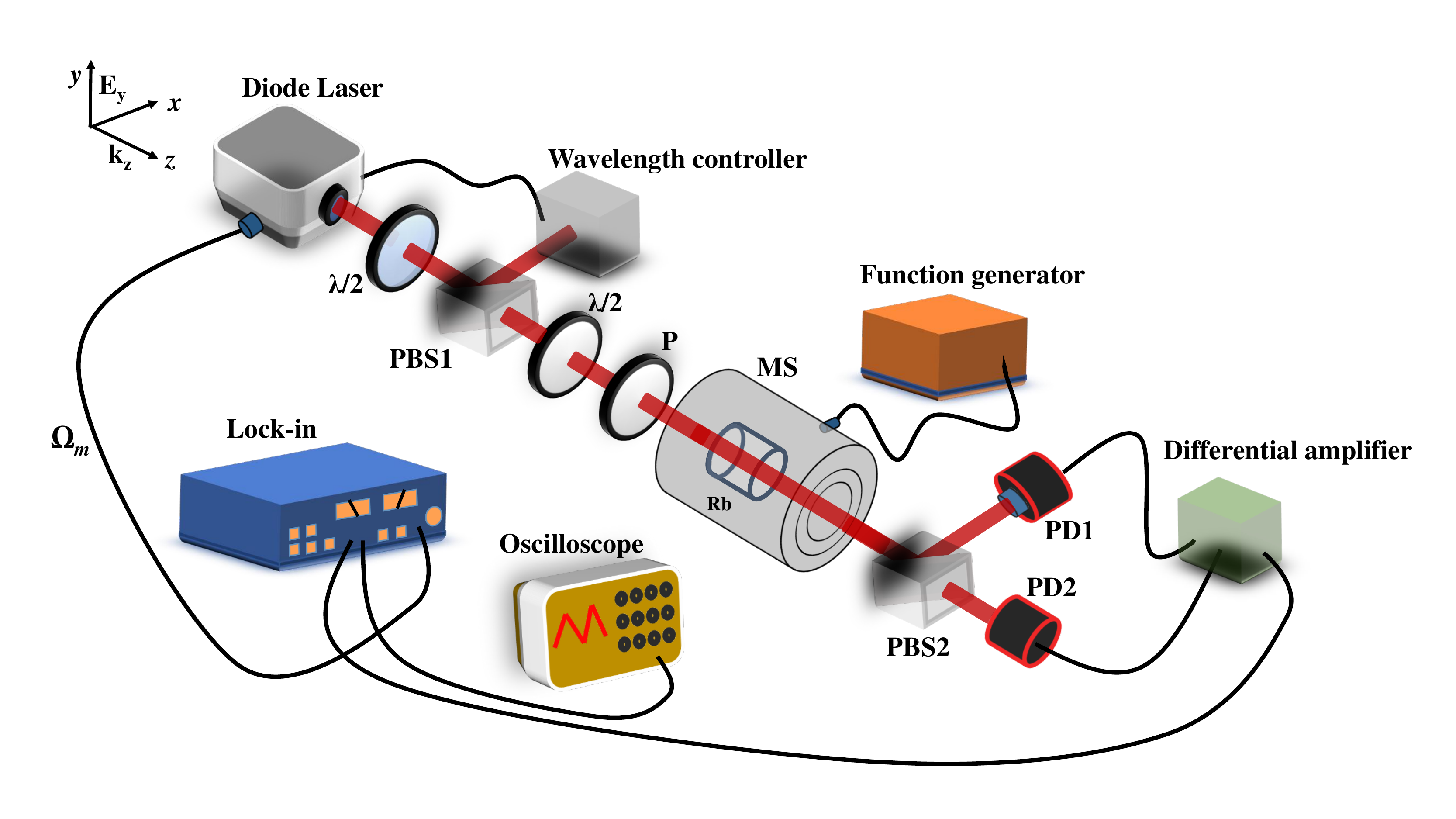}
\caption{ Schematic diagram of the experiment. PBS1 and PBS2 are the polarizing beam splitters; $\lambda /2$ is the half-wave plate; P is the polarizer; MS is the magnetic shield; PD1 and PD2 are the photo-detectors. The wavelength controller consists of the $\mu$DAVLL system and the wavemeter.}
\label{experimentalsetup}
\end{figure}

Figure~\ref{experimentalsetup} shows the schematic diagram of the experimental setup. A diode laser (Toptica DL pro), emitting light of a wavelength of 795~nm with a linewidth of less than 1 MHz, is used. The light frequency is modulated at $\Omega_{m}/2\pi=80$~kHz with a modulation depth of about 50~MHz to enable phase-sensitive detection and hence improve a signal-to-noise ratio of the observed signals.

 A half-wave plate $\lambda$/2 and a polarizing beam splitter PBS1, situated next to the laser, are used to control intensity of light directed into a dichroic atomic vapor laser lock (DAVLL), exploiting micro-machined vapor cell system \citep{Pustelny2016Dichroic}, and a wavemeter (HighFinesse/Angstrom Wavelength meter WS-U), used for wavelength stabilization and monitoring. The DAVLL setup is used to lock the light central frequency to the low-frequency wing of the Doppler-broadened $F=2\rightarrow F'=1$ transition of the $^{87}$Rb D$_1$ line.  The intensity of the light illuminating the vapors is set with the help of a half-wave plate $\lambda$/2 and a polarizer P. The main part of the light, polarized along $y$ direction, is sent through a paraffin-coated buffer-gas-free Rb vapor cell containing an isotopically enriched sample of $^{87}$Rb.  The vapor number density is controlled with the oven heating the cell to about 30$^\circ$C ($\approx 10^{10}$~atoms/cm$^3$).
The cell is placed at the center of a four-layer magnetic shield. The shield has three $\mu$-metal layers and one innermost ferrite layer (Twinleaf MS-1LF). The magnetic shield compensates the Earth magnetic field by a factor of 10$^6$ at center of the cell. To further reduce the magnetic field, the magnetic-field coils are installed inside the shield (not shown in Fig.~\ref{experimentalsetup}). These coils are also used to scan the magnetic field along the light propagation direction (the \textit{z}-axis) with the help of a function generator. The coils are also used to apply transverse magnetic field along the \textit{x}-axis using a computer controlled current source  (DM Technology Multichannel Current Source). In our experimental setup, since the residual-field compensation coils are  placed only along the \textit{x}-axis, a small residual field of the order of 14~$\mu$G is present inside the  shield along the \textit{y}-axis ($B_r=B_y\approx 14$~$\mu$G).
    
\begin{figure}[H]
\centering  
\includegraphics[height=8.2cm,width=\columnwidth]{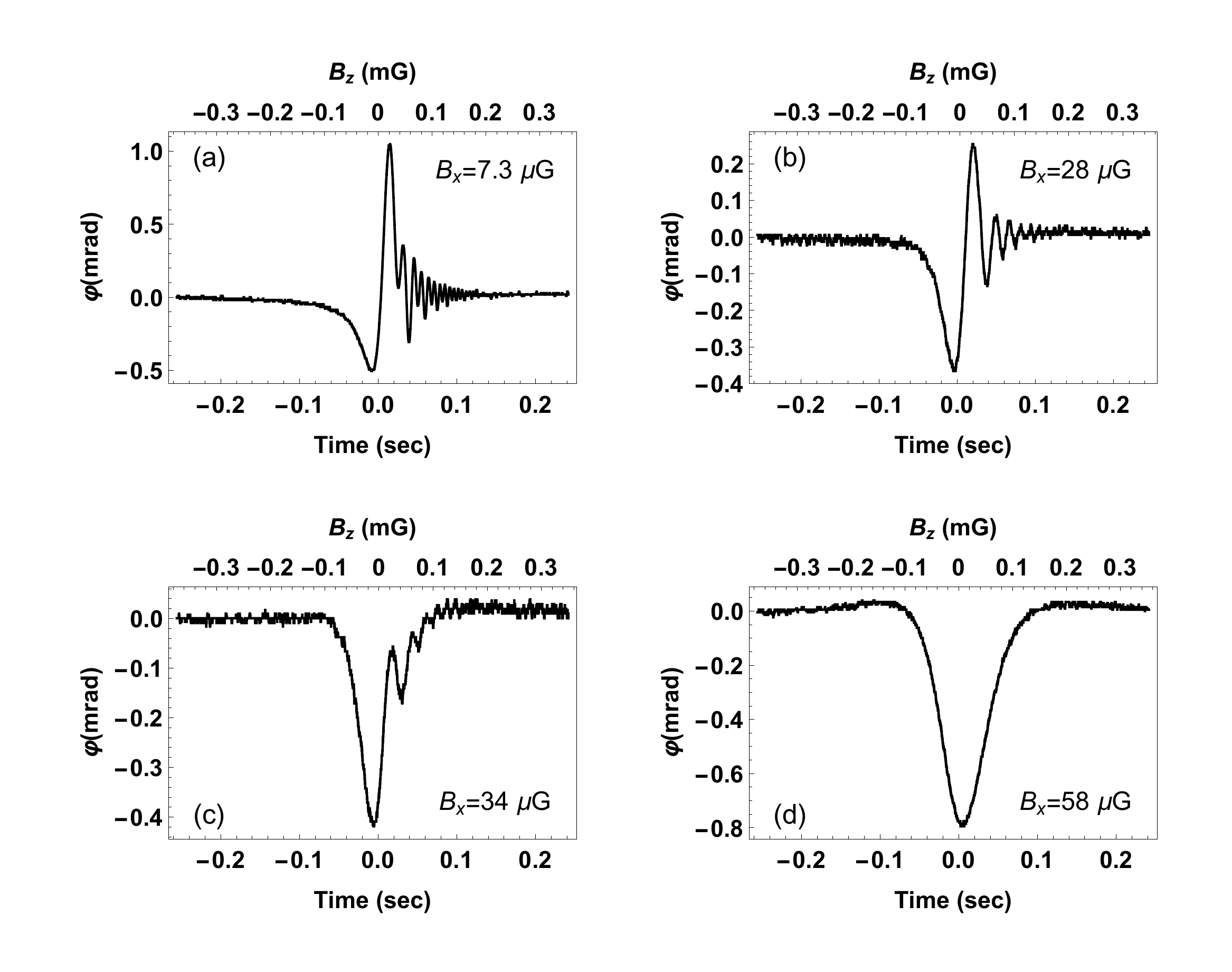}
\caption{Experimentally measured NMOR signals as a function of time/magnetic field at a sweep rate of 730 $\mu$G/s for the constant transverse magnetic field $B_y\approx 14$~$\mu$G and (a) $B_x= 7.3$~$\mu$G (b) $B_x$ = 28~$\mu$G (c) $B_x=34$~$\mu$G, and (d) $B_x=58$~$\mu$G.  The signals are measured with a light intensity of 2 $\mu$W/mm$^2$ before the cell.}
\label{ExpdiffTMFxy}
\end{figure}
The polarization rotation of light is measured using a balanced polarimeter setup, which contains a PBS2 and two photodiodes (PD1 and PD2). The PBS2 is rotated at 45$^{\circ}$ from the axis of the polarizer P and its outputs are measured using two photodiodes. The differential amplifier generates a difference signal which is demodulated with a lock-in amplifier at the first harmonic of the light modulation frequency. The time constant is small enough, not to affect the character of the observed dependences. Finally, the lock-in output signal is recorded with an oscilloscope.
     
Figure~\ref{ExpdiffTMFxy} shows the experimentally observed transient behavior  at different $B_x$  in the presence of residual magnetic along the \textit{y}-axis. The longitudinal magnetic field $B_z(t)$ is varied at a sweep rate of 730 $\rm{\mu G/s}$. The  results reveal a good agreement with theoretical calculations (Fig.~\ref{diffTMFxy}). Small transverse  fields modify the transient rotation signals and creates the two-frequency damped oscillations of polarization rotation [Fig.~\ref{ExpdiffTMFxy}(a)]. With a constant field ($B_y\approx 14$~$\mu$G) along the \textit{y}-axis, the oscillations at twice the Larmor frequency vanish and  the amplitude of the positive rotation decreases when $B_x$ is increased. Importantly, the strength of negative angles  remains almost unchanged [Fig.~\ref{ExpdiffTMFxy}(b)]. When $B_{x}$ increased further, no rotation at positive angle is observed [Fig.~\ref{ExpdiffTMFxy}(c)] and the shape of the signal becomes absorptive [Fig.~\ref{ExpdiffTMFxy}(d)]. This is a behavior that was also shown in Fig.~\ref{diffTMFxy}.

\begin{figure}[H]
\centering  
\includegraphics[height=4.30cm,width=\columnwidth]{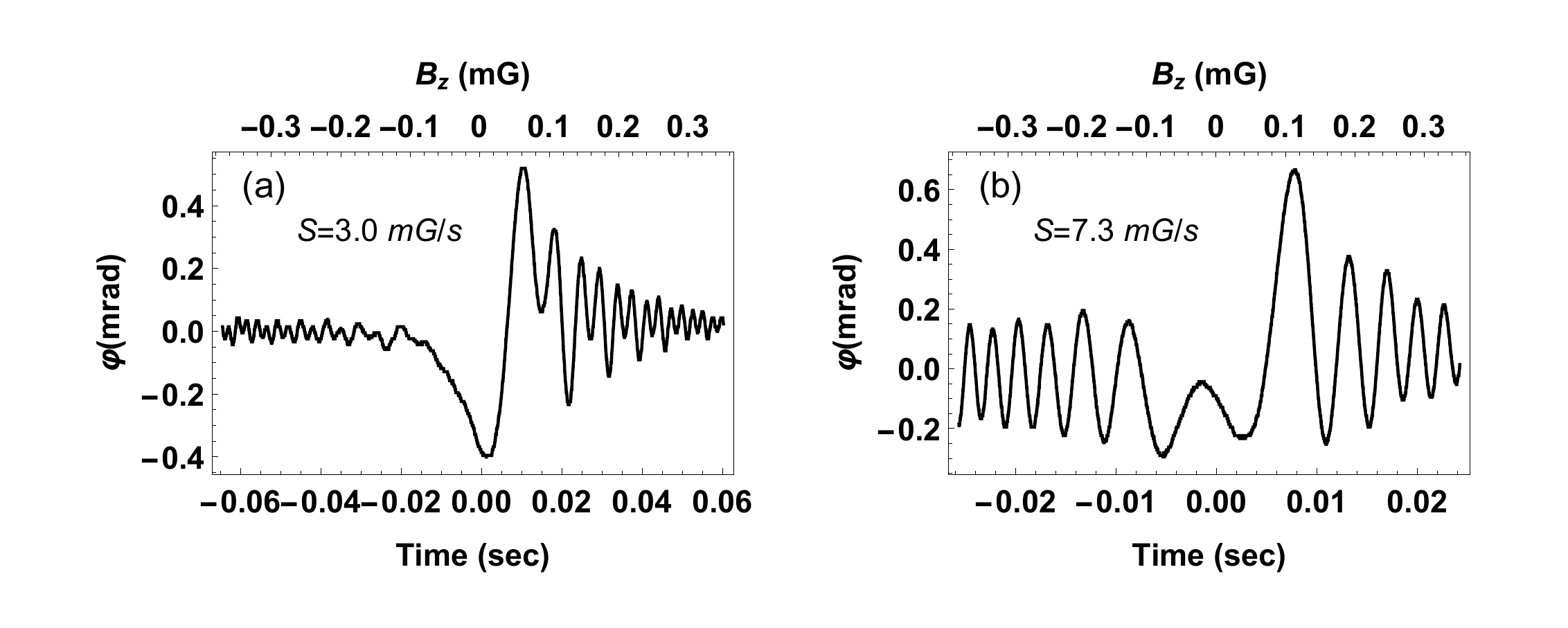}
\caption{Experimentally measured NMOR signals as functions of time/magnetic field at the sweep rate of (a) 3.0~mG/s and (b) 7.3~mG/s. The other parameters are same as in Fig.~\ref{ExpdiffTMFxy}(b). The oscillations observed at the left side of the resonance is the artifact of our measurements procedure \citep{RaghwinderNMOR}.}
\label{ExpdiffTMFDiffS}
\end{figure}

The residual magnetic field $B_y$ is very small, so it merely affects the signals at higher sweep rates in the absence of $B_x$. The effect of transverse fields at given values decreases when the sweep rate of signal is increased by keeping all other experimental parameters (light power, cell temperature etc.) same. For example, if the sweep rate in Fig.~\ref{ExpdiffTMFxy}(b) is increased, the oscillations at twice of the Larmor frequency reappears [Fig.~\ref{ExpdiffTMFDiffS}(a)]. As the sweep rate is further increased, the oscillations at the Larmor frequency disappears and the signal with a single frequency oscillations [Fig.~\ref{ExpdiffTMFDiffS}(b)], similar to that measured in the absence of the transverse fields, can be seen.
 
\section{Conclusion}

The transient dynamics of nonlinear magneto-optical rotation in a paraffin-coated Rb cell was studied in the presence of constant magnetic fields perpendicular to time-dependent longitudinal magnetic field  and light-propagation direction. Two-frequency oscillations in the transient polarization-rotation signals were observed for both transverse fields (parallel or perpendicular to the light polarization direction). As demonstrated, the behavior of the signals differs from  the steady-state NMOR signals, where the two-frequencies can be seen only when the transverse field is parallel to the light-polarization direction. We explained  the observations by using the angular probability surface of the density matrix of atoms. According to the explanation, when the transverse field is applied perpendicular to the light-polarization direction,  the atomic alignment tilts towards the light-propagation direction at an instantaneous longitudinal field close to zero. Hence, the angle between the atomic-alignment axis and the net magnetic field becomes less than 90$^{\circ}$  and the situation becomes similar to the case when the transverse field is parallel to light-polarization axis.  We demonstrated that the atomic alignment returns to its original state in the full Larmor period for both transverse fields. The amplitude of the oscillations at twice the Larmor frequency starts decreasing with increase of any transverse field as the angle between net magnetic field and alignment orientation gets smaller. For stronger fields, the rotation signals without oscillations are observed. When both transverse fields are non-zero at same time, the shape of the signals changes significantly. We observed the rotation signal with no transient oscillations and with minima near zero scanning field.

In this experiment, a small transverse field leads to the transient signals of different character than without the field. Therefore, by scanning the longitudinal field one may detect the transverse field in situ, and then compensate it by minimizing the $\Omega_L$-frequency component in the signal. With this respect, the effect can be used for precise compensation of the magnetic field and hence offers a possibility to extract full vectorial information of the magnetic field in a single measurement. Compensation of the field also enables decrease of the transverse relaxation rate of atoms.

Long relaxation times are also important for quantum-information storage, specific quantum-information-processing protocols (e.g., those employing optical dipole traps \citep{Weitenberg}) and even single-photon sources \citep{Wasilewski}. In this context, the ability to precise compensate transverse magnetic fields using a simple single-beam experiment may be particularly attractive. 

%\bibliographystyle{unsrt}%Choose a bibliograhpic style
%\nocite{*}
\bibliography{transienteffectTMF}
%\printbibliography

\end{document}